\documentclass[twocolumn,floatfix,superscriptaddress,a4paper,
               showpacs,showkeys,nofootinbib,reprint]{revtex4}
\usepackage{epsfig}
\usepackage{latexsym}
\usepackage{xspace}
\usepackage{hyperref}
\usepackage[latin2]{inputenc}
\usepackage{indentfirst}
\usepackage{enumerate}
\usepackage{color}

\usepackage{amsmath}
\usepackage{amssymb}
\usepackage[english]{babel}
\usepackage{url}
\newcommand{\eq}[1]{\begin{align} #1 \end{align}}

\begin{document}

\title{
Examination of the sensitivity of the thermal fits to heavy-ion hadron yield data to the modeling of the eigenvolume interactions
}
\author{Volodymyr Vovchenko}
\affiliation{
Frankfurt Institute for Advanced Studies, Goethe Universit\"at Frankfurt,
D-60438 Frankfurt am Main, Germany}
\affiliation{
Institut f\"ur Theoretische Physik,
Goethe Universit\"at Frankfurt, D-60438 Frankfurt am Main, Germany}
\affiliation{
Department of Physics, Taras Shevchenko National University of Kiev, 03022 Kiev, Ukraine}
\author{Horst Stoecker}
\affiliation{
Frankfurt Institute for Advanced Studies, Goethe Universit\"at Frankfurt,
D-60438 Frankfurt am Main, Germany}
\affiliation{
Institut f\"ur Theoretische Physik,
Goethe Universit\"at Frankfurt, D-60438 Frankfurt am Main, Germany}
\affiliation{
GSI Helmholtzzentrum f\"ur Schwerionenforschung GmbH, D-64291 Darmstadt, Germany}

\date{\today}

\begin{abstract}
The hadron-resonance gas (HRG) model with 
the
mass-proportional
eigenvolume (EV) corrections is employed to fit the hadron
yield data of the NA49 collaboration for central Pb+Pb collisions at $\sqrt{s_{_{\rm NN}}} = 6.3, 7.6, 8.8, 12.3,$ and $17.3$~GeV, the hadron midrapidity yield data of the STAR collaboration for Au+Au collisions at $\sqrt{s_{\rm NN}} = 200$~GeV, and the hadron midrapidity yield data of the ALICE collaboration for Pb+Pb collisions at $\sqrt{s_{\rm NN}} = 2760$~GeV.
At given bombarding energy, for a given set of radii, the EV HRG model fits do not just yield a single $T-\mu_B$ pair, but a whole range of $T-\mu_B$ pairs, each with similarly good fit quality. These pairs form a valley in the $T-\mu_B$ plane along a line of nearly constant entropy per baryon, $S/A$,
which increases nearly linearly 
with bombarding energy $E_{\rm lab}$.
The entropy per baryon values extracted from the data at the different energies are a robust observable: it is almost independent of the details of the modeling of the eigenvolume interactions and of the specific $T-\mu_B$ values obtained.
These results show that the extraction of the chemical freeze-out temperature and chemical potential is extremely sensitive to the modeling of the short-range repulsion between the hadrons. This implies that the ideal point-particle HRG values are not unique.
The wide range of the extracted $T$ and $\mu_B$ values suggested by the eigenvolume HRG fits, as well as the approximately constant $S/A$ at freeze-out, are consistent with a non-equilibrium scenario of continuous freeze-out, where hadrons can be chemically frozen-out throughout the extended space-time regions during the evolution of the system.
Even when the EV HRG fits are restricted to modest temperatures suggested by lattice QCD, the strong systematic effects of EV interactions are observed.
\end{abstract}



\maketitle

\section{Introduction}
The extraction of the thermodynamic properties of QCD is one of the major goals
of relativistic heavy-ion collision experiments. 
The phenomenological thermodynamic models are useful in this regard and have long been employed to estimate the temperatures reached in the relativistic heavy-ion collisions~\cite{SOG1981, Molitoris1985,Stoecker1986,Hahn1987}.
The thermal parameters at chemical freeze-out -- the stage of heavy-ion collision when inelastic reactions between hadrons cease -- have been extracted by fitting the rich data on hadron yields in various experiments,
ranging from the low energies at SchwerIonen-Synchrotron (SIS) to the highest energy of
the Large Hadron Collider
(LHC), within the hadron resonance gas (HRG) model~\cite{CleymansSatz,CleymansRedlich1998,CleymansRedlich1999,
Becattini2001,Becattini2004,ABS2006,ABS2009}.
It has been argued~\cite{DMB}, that the
inclusion of all known resonances into the model as free non-interacting (point-like) particles allows to
effectively model the attraction between hadrons.
Such formulation, a multi-component point-particle gas of all known hadrons and resonances, is presently the most commonly used one in the thermal model analysis.

HRG models which take into account the
repulsive interactions between hadrons are also available.
The HRG model with repulsive interactions have been compared with the lattice QCD data~\cite{EV-latt-1,EV-latt-2,EV-latt-3}, and
it has recently been shown in Ref.~\cite{VS2015} that the inclusion of the repulsive interactions into the HRG model in the form of a multi-component eigenvolume procedure can significantly change the values of chemical freeze-out temperature at $\mu_B=0$ while improving the agreement of the fit with the ALICE hadron yield data as compared to the point-particle HRG.
The analysis reveals a surprisingly strong sensitivity of the thermal fits at LHC to the modeling of eigenvolume interactions, leading to a large systematic uncertainty in the determination of chemical freeze-out temperature.
In the present work we perform a similar analysis at the finite (baryo)chemical potential by considering the data on hadron yields in Pb+Pb and Au+Au collisions of NA49 and STAR collaborations.
In order to study the sensitivity of the obtained results we use two different formulations of the multi-component eigenvolume HRG.

\section{Eigenvolume models}
The repulsive interactions between hadrons can be modeled by the eigenvolume correction of the van der Waals type, whereby the volume available for hadrons to move in is reduced by their eigenvolumes. Such a correction was included into the hadronic equation of state first in Refs.~\cite{HagedornEV1,Gorenstein1981,Kapusta1983}, while the thermodynamically consistent procedure for a single-component gas was formulated in Ref.~\cite{Rischke1991}.
In the present work we consider hadrons of different sizes, thus, a multi-component formulation of the eigenvolume HRG model is necessary. In our study we use two different formulations. Since both of these formulations are rigorously derived only for the case of Boltzmann statistics we will neglect the effects of quantum statistics 
in the present work. 

\subsection{``Diagonal'' eigenvolume model}
The single-component eigenvolume model of Ref.~\cite{Rischke1991} was generalized
to the multi-component case in Ref.~\cite{Yen1997}. It was assumed that the available volume of each of the hadron species is the same, and equals to the total volume minus the sum of the eigenvolumes of all hadrons in the system. 
Let us assume that we have $f$ different hadron species.
The pressure as function of the temperature and hadron densities has the following form
\eq{\label{eq:Pex1n}
P(T,n_1,\ldots, n_f) = T \sum_i \frac{n_i}{1 - \sum_j v_j n_j},
}
where the sum goes over all hadrons and resonances included in the model,
and where $v_i$ is the 
eigenvolume parameter of hadron species $i$.
The eigenvolume parameter $v_i$ can be identified with
the 2nd virial coefficient of the single-component gas of hard spheres and
is connected to the effective hard-core hadron radius as
$v_i = 4 \cdot 4 \pi r_i^3 / 3$.
In the grand canonical ensemble (GCE) one has to solve the non-linear equation for the pressure, which reads as
\eq{\label{eq:Pex1}
P(T, \mu) = \sum_{i} \, P^{\rm id}_i (T, \mu_i^*),
}
where $P^{\rm id}_i (T, \mu_i^*)$ is the pressure of the ideal (point-like) 
gas at the corresponding temperature and chemical potential, and
$\mu_i^* = \mu_i - v_i \, P(T, \mu)$ is the shifted chemical potential
due to the eigenvolume interactions. The $v_i$ is the eigenvolume parameter of the hadron species $i$, and the number density of these species can be calculated as
\eq{\label{eq:nex1}
n_i(T, \mu) = \frac{n_i^{\rm id} (T, \mu_i^*)}{1 + \sum_j v_j n_j^{\rm id} (T, \mu_j^*)}.
}

The multi-component eigenvolume HRG model given by Eqs.~\eqref{eq:Pex1n}-\eqref{eq:nex1} is the most commonly used one in the thermal model analysis.
Since this model does not describe properly the cross-terms in the virial expansion of the
multi-component gas of hard spheres (see details below) we will refer to it in this work as the ``Diagonal'' model.

\subsection{``Crossterms'' eigenvolume model}
The virial expansion of the classical (Boltzmann) multi-component gas of hard spheres up to 2nd order can be written as~\cite{LL}
\eq{\label{eq:virial}
P(T,n_1,\ldots, n_f) = T \sum_i n_i + T \sum_{ij} b_{ij} n_i n_j + \ldots,
}
where
\eq{\label{eq:bij}
b_{ij} = \frac{2 \pi}{3} \, (r_i + r_j)^3
}
are the components of the symmetric matrix of the 2nd virial coefficients.

Comparing Eqs.~\eqref{eq:Pex1n} and \eqref{eq:virial} one can see that the ``Diagonal'' model is not consistent with the virial expansion of the multi-component gas of hard spheres up to 2nd order and corresponds to a different matrix of 2nd virial coefficients, namely $b_{ij} = v_i$.
While we do not require hadrons to be non-deformable spherical objects and expect that the ``Diagonal'' model to capture the essential features of a system of particles with different sizes, the interpretation of $r_i$ as a hard-core hadron radius can be problematic in such model. Therefore, we additionally consider the 
multi-component eigenvolume model of Ref.~\cite{GKK}, which is formulated in the grand canonical ensemble (GCE) assuming Boltzmann statistics, and which is consistent with the 2nd order virial expansion in Eq.~\eqref{eq:virial}. The pressure in this model reads as
\eq{\label{eq:Pex2n}
P(T,n_1,\ldots, n_f) = \sum_i P_i = T \sum_i \frac{n_i}{1 - \sum_j \tilde{b}_{ji} n_j},
}
where
\eq{
\tilde{b}_{ij} = \frac{2\,b_{ii}\,b_{ij}}{b_{ii}+b_{jj}}
}
with $b_{ij}$ given by \eqref{eq:bij},
and where the quantities $P_i$ can be regarded as ``partial'' pressures.
This eigenvolume model given by \eqref{eq:Pex2n} is initially formulated in the canonical ensemble.
In Ref.~\cite{GKK} it was transformed to the grand canonical ensemble.
In the GCE formulation one has to solve the following system of non-linear 
equations for $P_i$
\eq{\label{eq:Pex2pi}
P_i = P_i^{\rm id} \left(T, \mu_i - \sum_j \tilde{b}_{ij} \, P_j \right), \qquad i = 1, \ldots , f,
}
where $f$ is the total number of the hadronic components in the model.
Hadronic densities $n_i$ can then be recovered by solving the system of linear equations connecting $n_i$ and $P_i$:
\eq{\label{eq:Pex2ni}
T n_i + P_i \sum_j \tilde{b}_{ji} n_j = P_i, \qquad i = 1, \ldots , f~.
}
We refer to the model given by Eqs.~\eqref{eq:Pex2n}-\eqref{eq:Pex2ni} as the ``Crossterms'' eigenvolume model. We note that this ``Crossterms'' model is very similar to the one used in Ref.~\cite{BugaevEV}.
From the technical point of view, the ``Crossterms'' model is more complicated than the ``Diagonal'' model: a set of coupled non-linear equations~\eqref{eq:Pex2pi} needs to be solved, instead of a single equation~\eqref{eq:Pex1} for the total pressure in the ``Diagonal'' model. In practice, the solution to \eqref{eq:Pex2pi} can be obtained by using an appropriate iterative procedure. In our calculations Broyden's method~\cite{Broyden} is employed to obtain the solution of the ``Crossterms'' model, using the corresponding solution of the ``Diagonal'' model as the initial guess.

\section{Calculation results}
\subsection{Some details about model implementation}
In our calculations we include strange and non-strange hadrons listed in the Particle Data Tables~\cite{pdg}, along with their decay branching ratios. This includes mesons up to $f_2(2340)$, (anti)baryons up to $N(2600)$.
We do not include hadrons with charm and bottom degrees of freedom which have a negligible
effect on the fit results, and we also removed the $\sigma$ meson ($f_0(500)$) and
the $\kappa$ meson ($K_0^*(800)$) from the particle list because
of the reasons explained in
Refs.~\cite{Broniowski:2015oha,Pelaez:2015qba}.
We also omit the light nuclei.
The finite width of the resonances is taken into account in the usual way, by adding the additional integration over their Breit-Wigner shapes in the point-particle gas expressions.
The feed-down from the decays of unstable resonances to the total hadron yields is included in the standard way. 

There are three conserved charges in the system: baryon charge, electric charge, and strangeness. Therefore, there are three corresponding independent chemical potentials: $\mu_B$, $\mu_Q$, and $\mu_S$. The chemical potential of the $i$th hadron species is thus determined as $\mu_i = B_i \mu_B + Q_i \mu_Q + S_i \mu_S$.
At each fixed temperature $T$ and baryochemical potential $\mu_B$, the $\mu_Q$ and $\mu_S$ are determined in a unique way in order to satisfy two conditions: the electric-to-baryon charge ratio of $Q/B = 0.4$, and the vanishing net strangeness. Assuming no pre-freezeout radiation, both of these conditions are relevant for the system created in the collision of heavy ions.

\subsection{Eigenvolume parametrizations}
As was mentioned before, the inclusion of the eigenvolume interactions is one of the most popular extensions of the standard HRG model.
In most of the analyses dealing with chemical freeze-out, which did include
the eigenvolume corrections~\cite{ABS2006,PHS1999,Cleymans2006}, it was assumed that all the hadrons have the same eigenvolume.
It has been established that, in this case, the eigenvolume corrections can significantly reduce the densities~\cite{Begun2013,mf-2014}, and, thus, increase the total system volume at the freeze-out as compared to the point-particle gas at the same temperature and same chemical potential. 
For this parametrization, however, the eigenvolume corrections essentially cancel out in the ratios of yields and, thus, have a negligible effect on the values of the extracted chemical freeze-out temperatures and chemical potentials.
If, however, one considers hadrons with different hard-core radii, then the ratios may change, and the fit quality can be improved~\cite{VS2015,Gorenstein1999,BugaevEV}. 

In our paper we use
a bag-model inspired parametrization of hadron eigenvolume. In this case the hadron eigenvolume is proportional to its mass through a bag-like constant, i.e. 
\eq{\label{eq:BagEV}
v_i = m_i / \varepsilon_0 .
}
This kind of eigenvolume parametrization had been obtained for the heavy Hagedorn resonances, and was used to describe their thermodynamical properties~\cite{HagedornEV1,Kapusta1983} as well as their
effect on particle yield ratios~\cite{NoronhaHostler2009}.
It was mentioned in the Ref.~\cite{Becattini2006} that such parametrization would lead to an increase of the freeze-out temperature, but that it does not entail an improvement of the fit quality in the ``Diagonal'' EV model, nor does it change other fit parameters. 
Note that the eigenvolume for the resonances with the finite width is assumed to be constant for each resonance, and is determined by its pole mass.

\begin{figure*}[!t]
\centering
\includegraphics[width=0.90\textwidth]{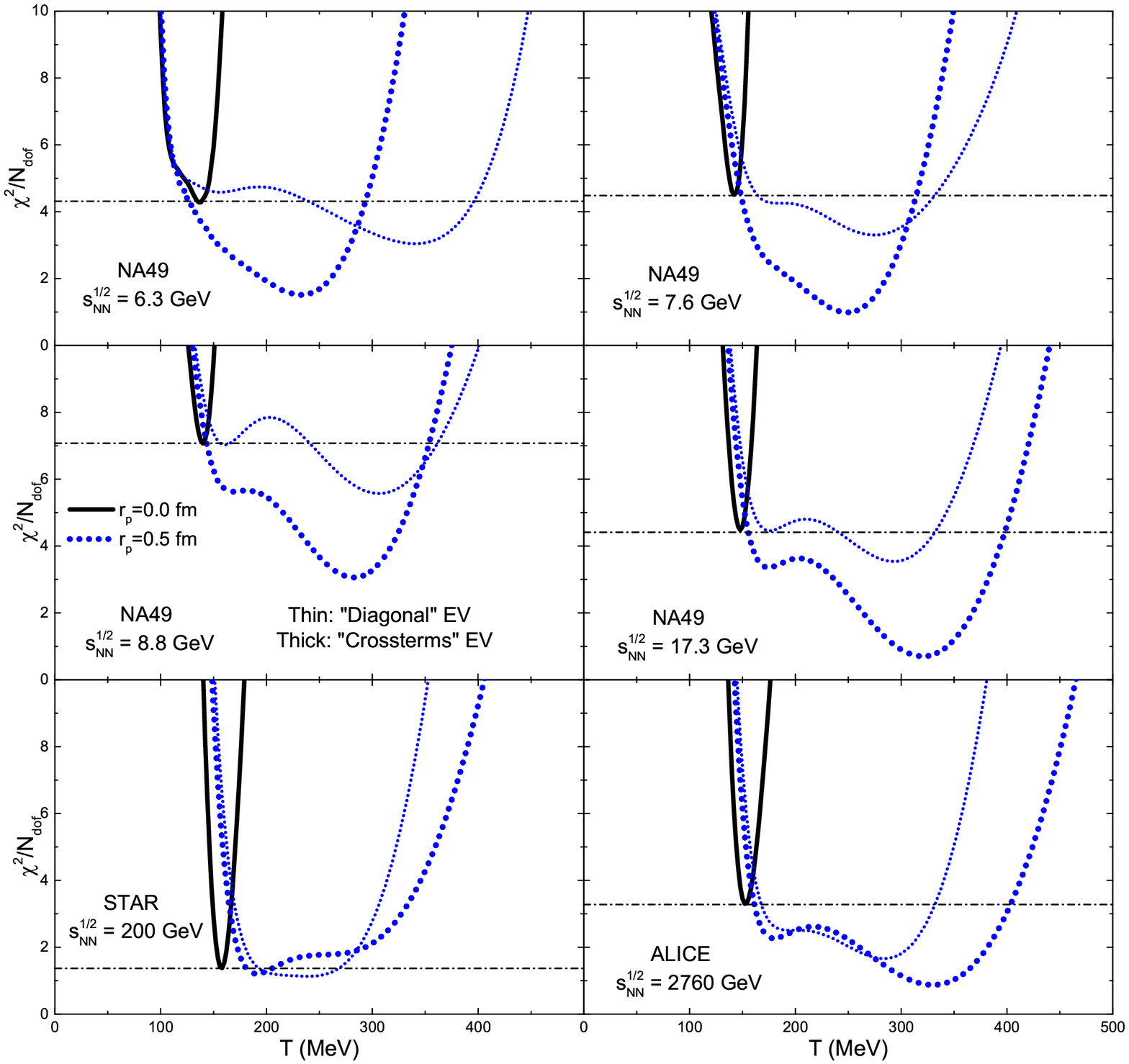}
\caption[]{(Color online)
The temperature dependence of $\chi^2 / N_{\rm dof}$ of fits to data of the NA49, STAR, and ALICE collaborations on hadron yields in central Pb+Pb and Au+Au collisions within the point-particle HRG model (solid black curves), the ``Diagonal'' eigenvolume model (thin dotted blue lines),
and the ``Crossterms'' eigenvolume model (thick dotted blue lines). 
The constant $\varepsilon_0$ is fixed in order to reproduce the effective hard-core proton radius of 
0.5 fm.
}\label{fig:chi2-vs-T}
\end{figure*}
\begin{figure}[!t]
\centering
\includegraphics[width=0.49\textwidth]{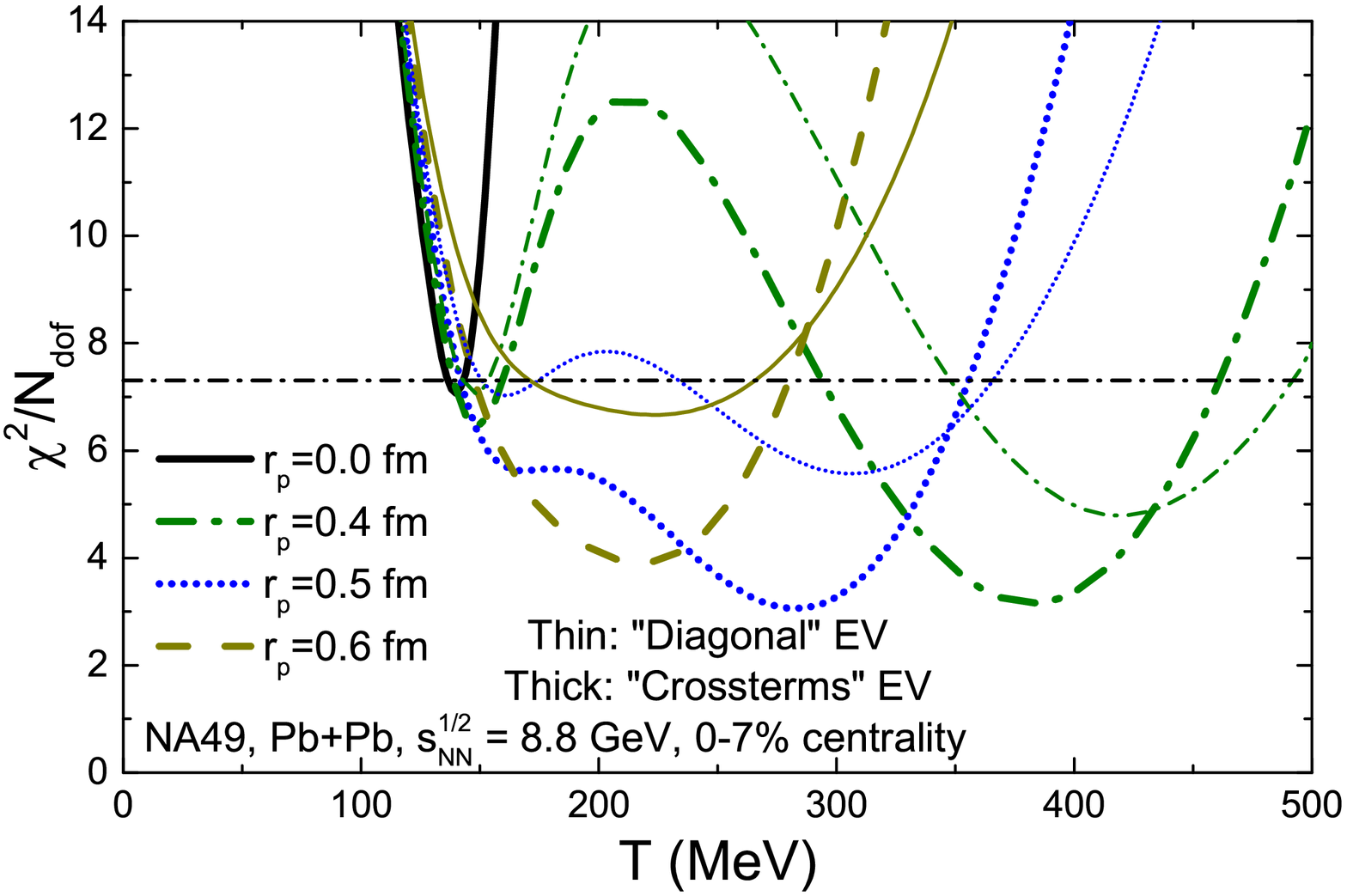}
\caption[]{(Color online)
The temperature dependence of $\chi^2 / N_{\rm dof}$ of fits to data of the NA49 Collaboration on hadron yields in central Pb+Pb collisions at $\sqrt{s_{_{\rm NN}}} = 8.8$~GeV within the point-particle HRG model (solid black curve), the ``Diagonal'' eigenvolume model (thin colored lines),
and the ``Crossterms'' eigenvolume model (differently styled colored lines). 
The constant $\varepsilon_0$ is fixed in order to reproduce the effective hard-core proton radius of 
0.4, 
0.5, and 0.6 fm.
}\label{fig:chi2-vs-T-40}
\end{figure}

Presently not much is well-established about the eigenvolumes of different hadron species, and  
it is debatable whether
parametrization \eqref{eq:BagEV} is the most realistic one. For instance, it can be argued, that strange hadrons should have a different (smaller) eigenvolume compared to non-strange ones. 
In general, the eigenvolumes might also depend on temperature and chemical potential.
In our study we would like to minimize the number
of the additional free parameters due to the introduction of the finite eigenvolumes and we use
\eqref{eq:BagEV} as a simple parametrization of hadron eigenvolumes.
The bag-like constant $\varepsilon_0$ determines the magnitude of the hadron eigenvolumes,
and in our analysis we vary this constant such that the corresponding hard-core radius $r_p$ of protons (nucleons) takes reasonable values.
Values of $r_p = 0.3$-$0.8$~fm have been rather commonly used in the literature~\cite{Yen1997,BugaevEV,Begun2013,PHS1999,Cleymans2006,Gorenstein1999}.
Additionally, the value $r_p \simeq 0.6$~fm was extracted from the ground state properties of nuclear matter within the fermionic van der Waals equation for nucleons~\cite{NM-VDW}. In the present work we vary the effective proton radius in the range $r_p = 0.0$-$0.6$~fm. Note that, within the bag-like parametrization, the radius $r_i$ of any hadron $i$ is related to the chosen value of $r_p$ through the relation $r_i = r_p \, \cdot \, (m_i/m_p)^{1/3}$, where $m_i$ is the mass of the hadron $i$.

\subsection{Experimental data set}

The thermal equilibrium EV HRG model fits are using the hadron yield
data of the NA49, STAR, and ALICE collaborations.

The data of the NA49 collaboration include
$4\pi$ yields of charged pions, charged kaons, $\Xi^-$, $\Xi^+$, $\Lambda$, $\phi$, and, where available, $\Omega$, $\bar{\Omega}$, measured in the 0-7\% most central Pb+Pb collisions at $\sqrt{s_{\rm NN}} = 6.3, 7.6, 8.8, 12.3$, and in the 0-5\% most central Pb+Pb collisions at $\sqrt{s_{\rm NN}} = 17.3$~GeV~\cite{NA49data-1,NA49data-2}. 
The feeddown from strong and electromagnetic decays is included in the model.
Additionally, the data on the total number of participants, $N_W$, is identified with total net baryon number and is included in the fit.
The actual tabulated data used in our analysis is available in Ref.~\cite{NA49data-3}.

The STAR data contains the midrapidity yields of charged pions, charged kaons,
(anti)protons, $\Xi^-$, $\Xi^+$, $\Omega$+$\bar{\Omega}$, and $\phi$ in the 0-5\% most central Au+Au collisions at $\sqrt{s_{\rm NN}} = 200$~GeV~\cite{STARdata,BecattiniSTAR}.
The yields of protons also include the feed-down from weak decays of (multi)strange hyperons, this is properly taken into account in the model. All other yields include the feeddown from strong and electromagnetic decays.

We also perform the fit to the ALICE data on 
midrapidity yields of hadrons in the 0-5\% most central Pb+Pb collisions at $\sqrt{s_{\rm NN}} = 2.76$~TeV~\cite{ALICEdata}. 
This includes yields of charged pions, charged kaons, (anti)protons, $\Xi^-$, $\Xi^+$, $\Lambda$, $\Omega$+$\bar{\Omega}$, $\phi$, and $K^0_S$.
Since the experimental centrality binning for $\Xi$ and $\Omega$ hyperons in the ALICE data is different from that of the other hadrons, we take the midrapidity yields of $\Xi$ and $\Omega$ in the $0-5$\% centrality class from Ref.~\cite{Becattini2014}, where they were obtained using the interpolation procedure.
All these yields include the feeddown from strong decays.

\subsection{Fitting results}

Figure~\ref{fig:chi2-vs-T} shows the temperature dependence of the $\chi^2/N_{\rm dof}$ of the fit to NA49, STAR, and ALICE data within ``Diagonal'' and ``Crossterms'' EV models for the value of the bag-like constant $\varepsilon_0$ fixed to reproduce the proton hard-core radius $r_p = 0.5$~fm. 
At each temperature the two remaining free parameters, namely the baryon chemical potential $\mu_B$ and the radius $R$ of the system volume $V = (4/3)\, \pi \, R^3$, are fitted in order to minimize the $\chi^2$ at this temperature. 

At this point, we do not enforce
any limitations on the values of $T$ and $\mu_B$ in our fitting procedure. This means, for instance, that we do not require that the eigenvolume HRG model describes the available lattice QCD data, and also that we employ no limitations on the packing fraction $\eta$ -- the fraction of the total volume occupied by hadrons of finite size. 
We will come back to these issues later.

The temperature dependence of the $\chi^2 / N_{\rm dof}$ within the point-particle HRG (solid black line in Fig.~\ref{fig:chi2-vs-T}) shows a narrow minimum, and the temperature of this minimum is slightly increasing with the collision energy. These temperature values are consistent with numerous previous analyses within the point-particle formulations of the HRG. The fit quality within the point-particle HRG is not very good. This is especially the case for the NA49 energy of $\sqrt{s_{\rm NN}} = 12.3$~GeV with $\chi^2 / N_{\rm dof} \simeq 13$. 
We note that the fit quality in the point-particle HRG can be improved significantly within the chemical non-equilibrium scenario (see e.g. Refs.~\cite{Becattini2006,Letessier2005,MHBQ,VBG2015}), however, we do not consider this option in the present work.

\begin{figure}[!t]
\centering
\includegraphics[width=0.49\textwidth]{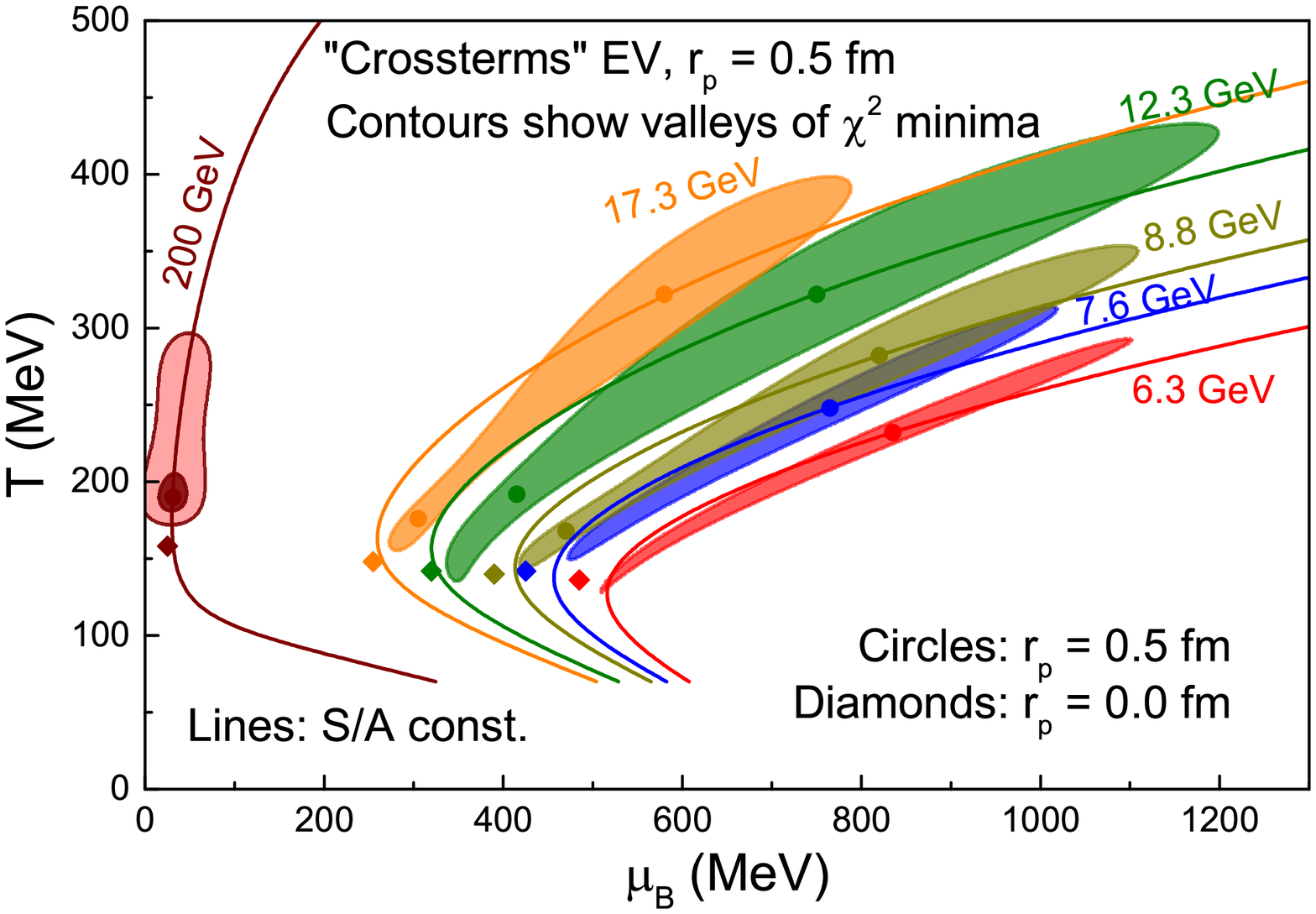}
\caption[]{(Color online)
Regions in the $T$-$\mu_B$ plane where the ``Crossterms'' eigenvolume HRG model with bag-like constant $\varepsilon_0$ fixed to reproduce the hard-core radius of $r_p = 0.5$~fm yields a better fit to the NA49 and STAR data as compared to the point-particle HRG model.
The locations of the local $\chi^2$ minima are depicted by the diamonds for the point-particle HRG case, and by the dots for the eigenvolume HRG. 
For RHIC energy the light shaded area additionally depicts the wide $T$-$\mu_B$ region where $\chi^2$ of the fit to the STAR data within the eigenvolume model has a broad minimum.
The solid lines show the isentropic curves for the eigenvolume model, which
go through the global $\chi^2$ minima.
}\label{fig:chi2-Tmu}
\end{figure}

\begin{table*}
 \caption{Summary of the fitted parameters within the point-particle HRG, and within the two eigenvolume HRG models with mass-proportional eigenvolumes fixed to $r_p = 0.5$~fm.
 No restrictions on possible values of $T$ and $\mu_B$ are applied.
 Full chemical equilibrium is assumed.} 
 \centering                                                 
 \begin{tabular*}{\textwidth}{c @{\extracolsep{\fill}} ccccc}                                   
 \hline
 \hline
 System & Parameters & Point-particle & ``Diagonal'' EV & ``Crossterms'' EV \\
  & & $r_p = 0$ & $r_p = 0.5$~fm & $r_p = 0.5$~fm \\
 \hline
 $\sqrt{s_{_{\rm NN}}} = 6.3$~GeV   & $T$ (MeV)                 
  & $138$ & $340$ & $233$ \\
  NA49                              & $\mu_B$ (MeV)             
  & $486$ & $1335$ & $838$ \\
  Pb+Pb                             & $R$ (fm)                  
  & $7.5$ & $7.2$  & $7.5$ \\
  0-7\% central 
  & $S/A$    
  & $13.2$        & $14.7$         & $14.4$        \\
  & $E/N$ (GeV)    
  & $1.11$        & $1.41$         & $1.25$        \\
  & $\chi^2 / N_{\rm dof}$    
  & $25.6/6$        & $18.3/6$         & $9.1/6$        \\
 \hline
 $\sqrt{s_{_{\rm NN}}} = 7.6$~GeV   & $T$ (MeV)                 
  & $142$ & $275$  & $249$ \\
  NA49                              & $\mu_B$ (MeV)             
  & $427$ & $862$ & $767$ \\
  Pb+Pb                             & $R$ (fm)                  
  & $7.8$ & $7.7$  & $7.7$ \\
  0-7\% central                     
  & $S/A$    
  & $15.4$        & $16.6$         & $17.3$        \\
  & $E/N$ (GeV)    
  & $1.09$        & $1.33$         & $1.25$        \\
  & $\chi^2 / N_{\rm dof}$    
  & $31.2/7$        & $23.1/7$         & $6.9/7$        \\
 \hline
 $\sqrt{s_{_{\rm NN}}} = 8.8$~GeV   & $T$ (MeV)                 
  & $140$ & $307$  & $283$ \\
  NA49                              & $\mu_B$ (MeV)             
  & $391$ & $903$  & $822$ \\
  Pb+Pb                             & $R$ (fm)                  
  & $8.8$ & $7.9$  & $8.3$ \\
  0-7\% central                     
  & $S/A$    
  & $17.7$        & $19.3$         & $19.8$        \\
  & $E/N$ (GeV)    
  & $1.03$        & $1.35$         & $1.28$        \\
  & $\chi^2 / N_{\rm dof}$    
  & $56.5/8$        & $44.6/8$         & $24.5/8$       \\
 \hline
 $\sqrt{s_{_{\rm NN}}} = 12.3$~GeV   & $T$ (MeV)                 
  & $142$  & $353$  & $322$ \\
  NA49                              & $\mu_B$ (MeV)             
  & $320$  & $898$  & $747$ \\
  Pb+Pb                             & $R$ (fm)                  
  & $10.0$ & $8.4$  & $8.3$ \\
  0-7\% central                     
  & $S/A$    
  & $23.1$        & $26.8$         & $27.4$        \\
  & $E/N$ (GeV)    
  & $0.97$        & $1.39$         & $1.32$        \\
  & $\chi^2 / N_{\rm dof}$    
  & $89.1/7$         & $64.2/7$         & $22.8/7$        \\
 \hline
 $\sqrt{s_{_{\rm NN}}} = 17.3$~GeV   & $T$ (MeV)                 
  & $148$  & $293$  & $320$ \\
  NA49                              & $\mu_B$ (MeV)             
  & $254$  & $534$  & $578$ \\
  Pb+Pb                             & $R$ (fm)                  
  & $10.4$ & $9.4$  & $9.1$ \\
  0-5\% central                     
  & $S/A$    
  & $29.3$        & $31.7$         & $35.0$        \\
  & $E/N$ (GeV)    
  & $0.98$        & $1.41$         & $1.27$        \\
  & $\chi^2 / N_{\rm dof}$    
  & $44.8/10$         & $35.4/10$         & $7.0/10$        \\
  \hline
 $\sqrt{s_{_{\rm NN}}} = 200$~GeV   & $T$ (MeV)                 
  & $157$  & $239$  & $190$ \\
  STAR                              & $\mu_B$ (MeV)             
  & $25$  & $38$  & $31$ \\
  Au+Au                             & $R$ (fm)                  
  & $8.1$ & $8.0$  & $8.7$ \\
  0-5\% central                     
  & $S/A$    
  & $300$        & $332$         & $324$        \\
  & $E/N$ (GeV)    
  & $0.97$        & $1.13$         & $1.01$        \\
  & $\chi^2 / N_{\rm dof}$    
  & $9.6/7$         & $7.9/7$         & $8.4/7$        \\
 \hline
 $\sqrt{s_{_{\rm NN}}} = 2760$~GeV   & $T$ (MeV)                 
  & $153$  & $284$  & $330$  \\
  ALICE                              & $\mu_B$ (MeV)             
  & 0 (fixed)        & 0 (fixed)        & 0 (fixed)        \\
  Pb+Pb                              & $R|_{y=0}$ (fm)                  
  & $11.0$ & $9.8$  & $9.3$ \\
  0-5\% central                      
  & $E/N$ (GeV)    
  & $0.93$        & $1.20$         & $1.28$        \\
  & $\chi^2 / N_{\rm dof}$    
  & $32.9/10$         & $16.6/10$         & $8.8/10$       \\
 \hline
 \end{tabular*}
\label{tab:FitUnconstrained}
\end{table*}

The fit quality within the two considered here eigenvolume models is considerably better than in the point-particle case at all energies, the best fits yield significantly higher values of the chemical freeze-out temperature. In Fig.~\ref{fig:chi2-vs-T-40} the temperature dependence of the $\chi^2$ for three different values of $r_p$, namely for $r_p = 0.4, \,0.5,\,0.6$~fm is shown for the SPS data at $\sqrt{s_{\rm NN}}=8.8$~GeV. The behavior of the $\chi^2$ for different values of $r_p$ shown in Fig.~\ref{fig:chi2-vs-T-40} is representative for all other collision energies considered in the present work.
For $r_p = 0.4$~fm the temperature dependence of $\chi^2$ shows two distinct local minima: the first one is located close to the minimum for point-particle HRG and the second one is at much higher temperatures and with considerably smaller $\chi^2$.
This trend is continued when even lower values of $r_p$ are considered.
For $r_p = 0.5$ fm and $r_p = 0.6$ fm a wide (double)minimum structure is observed in the temperature dependence of $\chi^2$ and, at all energies, the fit quality is better in EV models than in the point-particle case for a very wide high-temperature range. 
We note that the results presented in Fig.~\ref{fig:chi2-vs-T} for the ALICE energy are consistent with
the recent Ref.~\cite{VS2015}.
Small quantitative differences are attributed to the use
of the Boltzmann approximation in the present work.
The two eigenvolume models considered give the same qualitative picture, however there are significant quantitative differences, especially at higher temperatures: the ``Crossterms'' model consistently gives a considerably better description of the data.

In order to study the effects of the eigenvolume interactions on the fit values of the baryon chemical potential, we consider the structure of the $\chi^2$ of the fit in the $T$-$\mu_B$ plane.
For the purpose of clarity we consider just the ``Crossterms'' model with $r_p = 0.5$~fm.
Figure $\ref{fig:chi2-Tmu}$ depicts the regions in the $T$-$\mu_B$ plane where the fit quality
of NA49 data in the EV model is better than in the point-particle HRG.
At all five NA49 energies wide regions of improved $\chi^2$ values are observed at high temperatures and chemical potentials. We note a clear correlation between the fitted values of the chemical freeze-temperature and the chemical potential: the higher values of the temperature correspond to higher values of the baryochemical potential.
The thermal parameters at the global minima are listed in Table~\ref{tab:FitUnconstrained}. The values of the temperature at the global minima monotonically increase with the exception of the STAR energy. The corresponding values of the baryochemical potential monotonically decrease with an exception of a peak at $\sqrt{s_{\rm NN}} = 8.8$~GeV.

Based on the systematic analysis of the hadron yield data within the point-particle-like formulations of the HRG
it has been suggested that the chemical freeze-out is universally characterized by the constant average energy per hadron of $E/N \simeq 1$~GeV~\cite{CleymansRedlich1998}. 
It was shown in Ref.~\cite{Cleymans2006} that this criterium is robust with regards
to the eigenvolume corrections in the specific case when all hadrons have the same
hard-core radius.
It is interesting to analyze whether such a criterion holds as well for the bag-like eigenvolume parametrization considered in our work.
The values of $E/N$ at the global minimum for the point-particle, ``Diagonal'' EV, and  ``Crossterms'' EV with $r_p = 0.5$~fm are shown in the Table~\ref{tab:FitUnconstrained}.
The values of $E/N$ are notably larger when the eigenvolume corrections are present,
typically $E/N \simeq 1.3-1.4$~GeV.
Similar result, i.e. larger values of $E/N$, is observed for $r_p = 0.4$~fm and $r_p = 0.6$~fm. Therefore, one can conclude that the energy per particle criterium is, in general, \emph{not} robust with regards to the modeling of the eigenvolume interactions if the chemical freeze-out conditions are determined solely from the thermal fits.

\begin{figure}[!t]
\centering
\includegraphics[width=0.49\textwidth]{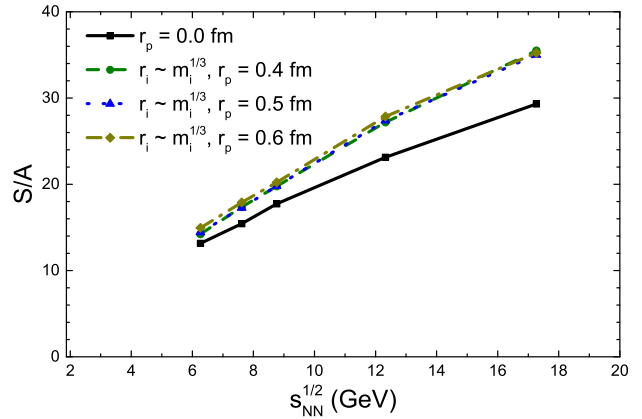}
\caption[]{(Color online)
The collision energy dependence of the entropy per baryon ratio $S/A$ calculated from the HRG model at the global minimum of the $\chi^2$ fit to the SPS hadron yield data. Calculations are done within the point-particle HRG and within the eigenvolume HRG with mass-proportional eigenvolumes. The parameter $\varepsilon_0$ in Eq.~\eqref{eq:BagEV} is fixed to reproduce the effective hard-core radius of proton of 0.4, 0.5, and 0.6 fm.
}\label{fig:SA}
\end{figure}

On the other hand, the total entropy per baryon $S/A$ is found to be much more robust.
This can be seen by looking at the $S/A$ values in Table~\ref{tab:FitUnconstrained}, on the lines of constant  $S/A$ in Fig.~\ref{fig:chi2-Tmu}, and on the 
excitation function of $S/A$ at SPS energies in Fig.~\ref{fig:SA}. 
It is seen that $S/A$ remains approximately constant across the vast $T$-$\mu_B$ islands of small $\chi^2$, and that $S/A$ increases only by 5-15\% when going from the point-particle HRG to the eigenvolume HRG, in spite of the massive changes in temperature and chemical potential.
It is also quite remarkable that the values of $S/A$ at the global minimum show almost no dependence on the value of $r_p$. This is best seen in Fig.~\ref{fig:SA} where the energy dependence of the $S/A$ values for $r_p = 0.4$, $0.5$, and $0.6$~fm extracted from- fits to the SPS data are shown.
These results hint towards the impossibility of fixing one ``best'' pair of temperature and chemical potential from a fit to the data: many $T-\mu_B$ pairs yield similarly good fits.
An isentropic expansion of matter with hadrons being chemically frozen out across the extended space-time regions is consistent with this finding. Namely, this implies that the chemical freeze-out is not a sharp process which takes place at one freeze-out hypersurface with similar values of the temperature and chemical potential. Rather a continuous process, occurring throughout the whole space-time evolution of the system created in the heavy-ion collisions, and characterized by a wealth of different values of temperatures, energy densities, and other dynamical parameters. Such a picture has been obtained within the transport model simulations of heavy-ion collisions by analyzing the space-time distribution of the chemical ``freeze-out'' points of various hadrons~\cite{ContFrz1,ContFrz2}.

In the present analysis we do not determine the uncertainties of the temperature and chemical potential extracted from fits to hadron yield data within the bag-like eigenvolume HRG model. These parameters depend strongly on the chosen eigenvolumes of the hadrons, and different parametrizations can give a fit of comparable quality but with very different values of $T$ and $\mu_B$.

\subsection{Fit constrained to low temperatures}

As seen from Figs.~\ref{fig:chi2-vs-T} and \ref{fig:chi2-Tmu} one can fit the data significantly better than in the point-particle case by employing the mass-proportional eigenvolume and considering high values of temperature and chemical potential.
However, the interpretation of these results is
prone to controversy: to describe QCD matter at $T>160$~MeV as an interacting
HRG with EV, i.e. with hadronic degrees of freedom, seems to contradict the consensual paradigm based on the ideal (point-particle) HRG model established in the community for decades.
Moreover, the high temperature $\chi^2$ minima shown in Fig.~\ref{fig:chi2-vs-T} are excellent fits of the model to the experimental data; however, the equation of state of this EV HRG model is plagued by the fact that 
the speed of sound 
takes values of $c_s^2 \sim 1$ at the $T$-$\mu_B$ values at the global minima for $r_p = 0.4-0.6$~fm.
The superluminal behavior of the speed of sound is a known problem of the EV model. Avoiding it requires modifications to the model.
Secondly, the packing fraction $\eta$ takes high values, typically 
$\eta \sim 0.15$
at the best fit locations.
At such high values of $\eta$ the eigenvolume model may deviate significantly from the equation of state of the hard spheres model~(see, e.g., Refs.~\cite{mf-2014,ZalewskiRedlich}). Evidently, this suggests that the global minima
locations may lie outside the range where the model can be applied safely.
In this respect it would be interesting to study the behavior of eigenvolume HRG 
at high densities within other EV formulations which take care of these issues.
Such a study is outside the scope of the present paper.

The mass-proportional eigenvolume interactions give a systematic improvement of the fit quality at all considered collision energies. 
However, the interpretation of these results is evidently controversial, for the reasons listed above.
Thus, we also perform an additional calculation where we restrict the temperatures from above, in order to get some consistency with the lattice QCD data.

The lattice QCD calculations reveal a crossover-type transition between hadronic and partonic degrees at $\mu_B = 0$, with a pseudocritical temperature $T_{\rm pc} \sim 155$~MeV\cite{Aoki:2006we,lQCD,Bazavov:2014pvz}.
The lattice data, however, does not give a direct information about the constituent composition of matter at a particular temperature. Due to the crossover nature of the transition it is impossible to uniquely define a transition temperature, which would exclude a presence of the hadronic degrees above this temperature. For example, it was illustrated in~\cite{Mukherjee:2015mxc} that the charm hadron-like excitations remain dominant degrees of freedom at temperatures above the pseudocritical one, up to at least $1.2 \, T_{\rm pc}$~(see also \cite{Biro:2014sfa})
while the $T_{\rm pc} \sim 155$~MeV temperature suggests only the onset of hadron melting~\cite{Jakovac:2013iua}.
Such a picture, a \emph{gradual} transition from hadrons to quarks had been advocated before, in particular in the context of heavy-ion collisions~\cite{Stoecker:1980uk}.

The thermodynamics of the HRG with mass-proportional eigenvolumes was considered in Ref.~\cite{EV-latt-3} in the
context of the lattice QCD equation of state, in particular regarding the gradual transition from hadrons to quarks.
In that work a crossover equation of state of the QCD matter was developed. It was obtained by smoothly matching two models, an 
eigenvolume HRG equation of state
at low $T$ and/or $\mu_B$ and a perturbative QCD (pQCD) equation of state at high $T$ and/or $\mu_B$. 
The total QCD pressure function at vanishing chemical potential in this crossover model reads
\eq{
P(T) = S(T) \, P_{\rm qg} (T) + [1 - S(T)] \, P_{\rm h} (T),
}
where $P_{\rm h} (T)$ is the hadronic pressure modeled by the eigenvolume HRG,
$P_{\rm qg} (T)$ is the pressure of the quark-gluon phase based on the pQCD calculations, and
\eq{
S(T) = \exp\left[-(T_0/T)^{r} \right]
}
is the so-called switching function.
In one of the scenarios considered in Ref.~\cite{EV-latt-3} the hadronic part was modeled by the ``Diagonal'' eigenvolume HRG model with the mass-proportional eigenvolume \eqref{eq:BagEV}. The bag-like constant $\varepsilon_0$ was determined from the fit to the lattice data. The best fit was obtained for $\varepsilon_0 = 797$~MeV/fm$^3$, which in our notation corresponds to the hard-core proton (nucleon) radius of $r_p \simeq 0.41$~fm, and a ``switching'' temperature of $T_0 \sim 175$~MeV.
Using the point-particle HRG gives a worse but still a satisfactory fit to the lattice data at $\mu_B = 0$, thus, it seems that values of hard-core proton radius in the range $r_p = 0.00-0.41$~fm are all generally compatible with the lattice data within the mass-proportional EV model.

\begin{figure}[t]
\centering
\includegraphics[width=0.49\textwidth]{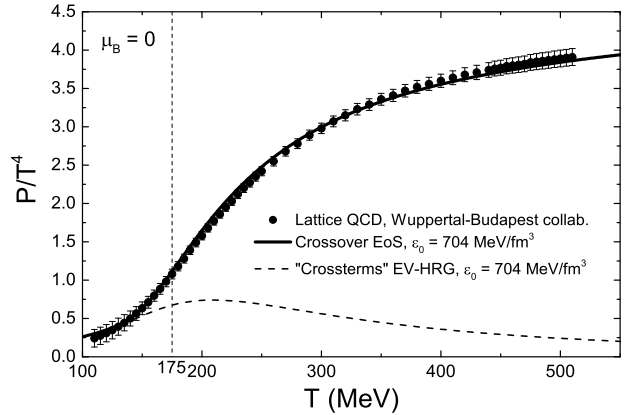}
\caption[]{(Color online)
Temperature dependence of the scaled pressure at zero baryon chemical potential calculated within the crossover equation of state (solid line), and
the ``Crossterms'' eigenvolume HRG with $v_i = m_i / \varepsilon_0$ and $\varepsilon_0 = 704$~MeV/fm$^3$~(dashed line).
The lattice QCD data of the Wuppertal-Budapest collaboration~\cite{lQCD} is shown by the symbols with error bars.
}\label{fig:crossoverEoS}
\end{figure}

\begin{table*}
 \caption{Summary of the fitted parameters within the HRG models restricted to temperatures below 175~MeV. Chemical equilibrium is assumed.} 
 \centering                                                 
 \begin{tabular*}{\textwidth}{c @{\extracolsep{\fill}} cccc}                                   
 \hline
 \hline
 System & Parameters & Point-particle  & ``Crossterms'' EV \\
  & & $r = 0$~fm & $r_i \sim m_i^{1/3}$, $r_p = 0.43$~fm \\
 \hline
 $\sqrt{s_{_{\rm NN}}} = 6.3$~GeV   & $T$ (MeV)                 
  & $137.5 \pm 5.3$ & $158.6 \pm 14.1$ \\
  NA49                              & $\mu_B$ (MeV)             
  & $485.7 \pm 4.1$ & $555.2 \pm 36.5$ \\
  Pb+Pb                             & $R$ (fm)          
  & $7.46  \pm 0.70$ & $7.71 \pm 0.59$  \\
  0-7\% central                     & $\chi^2 / N_{\rm dof}$   
  & $25.6/6$        & $20.2/6$         \\
 \hline
 $\sqrt{s_{_{\rm NN}}} = 7.6$~GeV   & $T$ (MeV)                 
  & $142.4 \pm 3.7$ & $159.4 \pm 5.9$ \\
  NA49                              & $\mu_B$ (MeV)             
  & $427.2 \pm 3.7$ & $479.5 \pm 12.9$ \\
  Pb+Pb                             & $R$ (fm)                  
  & $7.82 \pm 0.41$ & $8.29 \pm 0.30$  \\
  0-7\% central                     & $\chi^2 / N_{\rm dof}$   
  & $31.2/7$        & $23.9/7$         \\
 \hline
 $\sqrt{s_{_{\rm NN}}} = 8.8$~GeV   & $T$ (MeV)                 
  & $140.0 \pm 2.5$ & $151.6 \pm 4.5$  \\
  NA49                              & $\mu_B$ (MeV)             
  & $391.2 \pm 5.0$ & $425.1 \pm 9.8$ \\
  Pb+Pb                             & $R$ (fm)                  
  & $8.83 \pm 0.42$ & $9.21 \pm 0.37$  \\
  0-7\% central                     & $\chi^2 / N_{\rm dof}$   
  & $56.2/8$        & $50.4/8$         \\
 \hline
 $\sqrt{s_{_{\rm NN}}} = 12.3$~GeV   & $T$ (MeV)                 
  & $141.7 \pm 3.1$  & $157.0 \pm 8.0$  \\
  NA49                              & $\mu_B$ (MeV)             
  & $320.2 \pm 4.9$  & $347.1 \pm 9.3$ \\
  Pb+Pb                             & $R$ (fm)                  
  & $10.00 \pm 0.54$  & $9.90 \pm 0.64$  \\
  0-7\% central                     & $\chi^2 / N_{\rm dof}$   
  & $89.1/7$         & $80.4/7$         \\
 \hline
 $\sqrt{s_{_{\rm NN}}} = 17.3$~GeV   & $T$ (MeV)                 
  & $147.9 \pm 2.1$  & $160.2 \pm 3.9$  \\
  NA49                              & $\mu_B$ (MeV)             
  & $254.5 \pm 4.7$  & $277.4 \pm 6.6$ \\
  Pb+Pb                             & $R$ (fm)                  
  & $10.44 \pm 0.41$ & $10.89 \pm 0.37$  \\
  
  0-5\% central                     & $\chi^2 / N_{\rm dof}$   
  & $44.8/10$         & $38.7/10$         \\
 \hline
 $\sqrt{s_{_{\rm NN}}} = 200$~GeV   & $T$ (MeV)                 
  & $157.4 \pm 2.3$  & $173.4 \pm 5.1$  \\
  STAR                              & $\mu_B$ (MeV)             
  & $25.3 \pm 11.2$  & $27.9 \pm 12.7$  \\
  Au+Au                             & $R|_{y=0}$ (fm)                  
  & $8.13 \pm 0.33$  & $8.55 \pm 0.32$  \\
  
  0-5\% central                     & $\chi^2 / N_{\rm dof}$    
  & $9.6/7$         & $8.6/7$          \\
 \hline
 $\sqrt{s_{_{\rm NN}}} = 2760$~GeV   & $T$ (MeV)                 
  & $152.9 \pm 2.3$  & $165.8 \pm 4.3$  \\
  ALICE                              & $\mu_B$ (MeV)             
  & 0 (fixed)        & 0 (fixed)        \\
  Pb+Pb                              & $R|_{y=0}$ (fm)                  
  & $10.98 \pm 0.47$ & $11.35 \pm 0.45$ \\
  
  0-5\% central                      & $\chi^2 / N_{\rm dof}$    
  & $32.9/10$         & $27.0/10$         \\
 \hline
 \end{tabular*}
\label{tab:FitCo}
\end{table*}

Our hadron list is very similar to the one used in Ref.~\cite{EV-latt-3}. 
On the other hand, we believe that the ``Crossterms'' model reflects better the features of the multi-component eigenvolume systems. In particular, as previously mentioned, it is consistent with virial expansion of classical hard spheres equation of state, in contrast to ``Diagonal'' model.
As seen from Fig.~\ref{fig:chi2-vs-T} it also gives a systematically better description of the hadron yield data.
We use the ``Crossterms'' EV model and repeat the calculation of Ref.~\cite{EV-latt-3} for the crossover equation of state. For the pQCD part we use exactly the same model with exactly the same parameters. They are listed in the second-last row of Table~1 in Ref.~\cite{EV-latt-3}. 
We reproduce the result of~\cite{EV-latt-3}, i.e. a perfect description of the lattice QCD pressure, by using the value
$r_p \simeq 0.43$~fm ($\varepsilon_0 = 704$~MeV/fm$^3$) 
for the proton hard-core radius
within the ``Crossterms'' EV model.
The slightly different value of $\varepsilon_0$ compared to~\cite{EV-latt-3} is attributed to differences between the ``Diagonal'' and ``Crossterms'' models.
The resulting temperature dependence of the scaled pressure $P/T^4$ is shown
in Fig.~\ref{fig:crossoverEoS}. A very good agreement is seen in the whole temperature range, in line with results of Ref.~\cite{EV-latt-3}. 
The dashed line in Fig.~\ref{fig:crossoverEoS} shows the purely hadronic pressure $P_h$ of the eigenvolume HRG.
The bag-like eigenvolume models with $r_p = 0.4-0.6$~fm show a qualitatively similar temperature dependence, with larger $r_p$ values bringing the dashed curve further down.

Let us now employ the
``Crossterms'' model with $r_i \sim m_i^{1/3}$ and $r_p = 0.43$~fm to fit the NA49, STAR, and ALICE data, and impose an additional restriction $T < T_0 \simeq 175$~MeV. This ensures that only the temperatures where the hadronic part of the QCD equation of state plays a notable role are considered. 
Certainly, due to the crossover nature of the hadron-parton transition, this value should not and cannot be regarded as a sharp transition temperature.

\begin{figure}[t]
\centering
\includegraphics[width=0.49\textwidth]{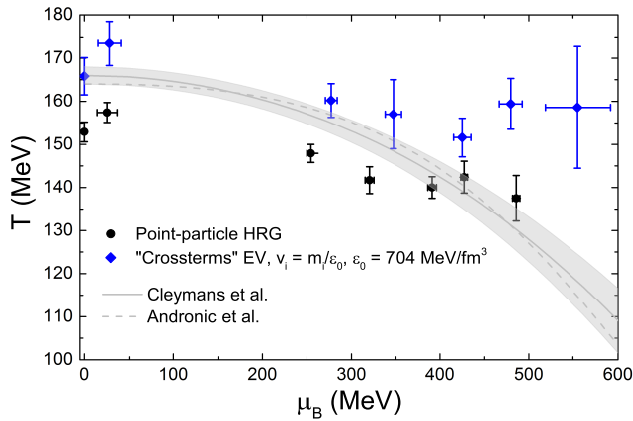}
\caption[]{(Color online)
The extracted freeze-out parameters within the point-particle HRG,
and the case when eigenvolumes of hadrons are proportional to their mass with $\varepsilon_0 = 704$~MeV/fm$^3$, modeled
by the ``Crossterms'' eigenvolume HRG.
The parameterized freeze-out curves from Refs.~\cite{ABS2009}, \cite{Cleymans2006}, 
obtained within the point-particle-like (including constant eigenvolume for all hadrons) HRG models, are depicted by lines.
}\label{fig:FitCo}
\end{figure}

Even with the low temperature restriction, the EV model fits leads to a better description of the data at all the considered energies, as seen in Table~\ref{tab:FitCo}. The improvement is modest but systematic.
The inclusion of the finite EV also leads to some changes in the extracted parameters: the chemical freeze-out temperature increases by about 10-15 MeV and
the baryochemical potential increases by about 10-15\%.
These changes are significant as they are larger than the typical
thermal fit uncertainties reported in the literature.
The resulting temperature profiles of the $\chi^2$
are all qualitatively similar to the $r_p = 0.4$~fm curve in Fig.~\ref{fig:chi2-vs-T-40}.
The fit errors of $T$ and $\mu_B$, 
obtained from analyses of the second-derivative error matrices at the minima,
increase notably for the cases with finite EV.
The obtained results also indicate that the chemical freeze-out curve in $T$-$\mu_B$ plane has a smaller curvature in the EV models compared to the one obtained within the point-particle HRG~(see Fig.~\ref{fig:FitCo}). A similar result was obtained in \cite{Becattini2013,Becattini:2016xct} but by employing a different mechanism, namely, by considering the distortion of yields due to the post-hadronization cascade phase. 
Due to the possibly irregular structure of the $\chi^2$ profiles one should treat the uncertainties for the extracted freeze-out parameters shown in Table~\ref{tab:FitCo} and Fig.~\ref{fig:FitCo} with care,
since they are calculated from the second-derivative error matrix and
are based on an assumed parabolic behavior of $\chi^2$ near the minima.

It is seen that the extraction of the chemical freeze-out parameters is rather sensitive to the modeling of repulsive interactions between hadrons even if only moderate temperatures $T<175$~MeV are considered.
As mentioned before, the lattice data for the pressure does not conclusively exclude any $r_p$ in the range of $0.00-0.43$~fm for the bag-like parametrization of the hadron eigenvolumes. It is also likely that the upper limit of acceptable $r_p$ is even higher. For these reasons the uncertainties in the extraction of the chemical freeze-out parameters remain sizable even when the lattice constraint is used.

\section{Summary}

In summary, the data of the NA49, STAR, and ALICE collaborations on the hadron yields in central Pb+Pb (Au+Au) collisions at $\sqrt{s_{\rm NN}} = 6.3, 7.6, 8.8, 12.3, 17.3$, $200$, and $2760$~GeV are analyzed within two different multi-component eigenvolume HRG models employing 
mass-proportional eigenvolumes
for different hadrons.
For a proton hard-core radius of 0.4-0.6~fm, these models describe the data significantly better than the conventional point-particle HRG model in very wide regions in the $T$-$\mu_B$ plane.

These results show that the extraction of the chemical freeze-out parameters is extremely sensitive to the modeling of the short-range repulsion between the hadrons, and they imply that the ideal point-particle HRG values are not unique.
As far as the thermal fits within the EV-HRG model are concerned, 
the constant energy per hadron of 1 GeV criterion proposed in the literature is 
changing ($\sim$30\%) depending on 
the modeling of the eigenvolume interactions.
On the other hand,
the entropy per baryon extracted from the data for the different energies is found to be much more robust: it is nearly independent of the details of modeling of the eigenvolume interactions and of the specific $T-\mu_B$ values obtained.
This, as well as the
large uncertainties in the extracted $T$ and $\mu_B$ values suggested by the eigenvolume HRG models, is consistent with the scenario of a continuous freeze-out, whereby the hadrons are being frozen-out throughout the extended regions of the space-time evolution of the system rather than from the sharp freeze-out hypersurface.

The interpretation of the high values of temperature and baryochemical potential obtained within EV fits is prone to controversy. These values do give a significant and systematic improvement in the fit quality. However, the fits are also plagued by an irregular behavior of the speed of sound, difficulty in reconciliation with lattice QCD results, and high values of packing fraction. Thus, the extracted high values of $T$ and $\mu_B$ should not be interpreted as estimates of chemical freeze-out conditions, but rather as an illustration of the hitherto unexplored sensitivity of thermal fits to the modeling of EV interactions. 
Even the fits with a low temperature ($T<175$~MeV) restriction inspired by analysis of the lattice QCD equation of state suggest a strong influence of EV effects on thermal fits. The fits with this temperature restriction result into 10-15~MeV higher values of the chemical freeze-out temperature and a systematically improved $\chi^2$
compared to the point-particle case.

Hence, the modeling of the eigenvolume interactions plays 
a crucial role for the thermal fits to the hadron yield data, a fact that had been largely overlooked in the past.
Many possibilities for the parameterization of the hadron eigenvolumes exist.
It is evident that proper restrictions on the eigenvolumes of different hadrons are urgently needed. 
Any conclusions based on thermal fits must be based not just on the location of the $\chi^2$ minimum and its magnitude, but rather on the full profile of the $\chi^2$. 
The $\chi^2$ may have 
non-parabolic structure around the local minima, and thus, the standard statistical-based estimates of the uncertainties of the extracted parameters may become inapplicable.

The analysis is performed within two different multi-component eigenvolume HRG models: the ``Diagonal'' EV model  and the ``Crossterms'' EV model. Both models yield qualitatively similar results but they differ quantitatively. 
The commonly used ``Diagonal'' EV model is shown to be not consistent with the 2nd order virial expansion for the equation of state of the multi-component system of hard spheres, while technically more complicated ``Crossterms'' EV model is. 
The ``Crossterms'' model also opens new possibilities
to model the repulsive interactions: it allows to
independently specify the virial coefficient $b_{\rm ij}$ 
between any pair of hadron species.
For example, one can treat differently the baryon-baryon, baryon-antibaryon, meson-baryon, and meson-meson repulsive interactions. These effects were recently explored for LHC energies in Ref.~\cite{Satarov:2016peb}.
Thus, the ``Crossterms'' model is suggested to be used 
over the ``Diagonal'' one
in the future
analyses employing the EV corrections.

The collision energy range investigated in this work is relevant for the ongoing SPS and RHIC beam energy scan programs, as well as for the experiments at the future FAIR and NICA facilities.
The strong effects of EV interactions should be taken into account in the future analyses and interpretations of the hadron yield data.

\vspace{0.5cm}
\section*{Acknowledgements}
We are grateful
to
P.~Alba, 
M.~Ga\'zdzicki,
M.I.~Gorenstein, L.M.~Satarov, and A.~Tawfik
for fruitful comments and discussions.
This work was supported by the Helmholtz International Center for FAIR within the LOEWE program of the State of Hesse.
Some of the numerical calculations were performed at the Prometheus cluster at GSI.
V.V. acknowledges the support from HGS-HIRe for FAIR.
H.St. appreciates the support through the Judah M.~Eisenberg Laureatus Chair
at Goethe University.


\begin{thebibliography}{9}
\bibitem{SOG1981}
H.~Stoecker, A.~A.~Ogloblin, and W.~Greiner, Z. Phys. A 303, 259 (1981).

\bibitem{Molitoris1985} 
  J.~J.~Molitoris, D.~Hahn, and H.~Stoecker,
  Prog.\ Part.\ Nucl.\ Phys.\  {\bf 15}, 239 (1985).

\bibitem{Stoecker1986} 
  H.~Stoecker and W.~Greiner,
  Phys.\ Rept.\  {\bf 137}, 277 (1986).

\bibitem{Hahn1987} 
  D.~Hahn and H.~Stoecker,
  Phys.\ Rev.\ C {\bf 35}, 1311 (1987).

\bibitem{CleymansSatz}
J. Cleymans and H. Satz, Z. Phys. C 57, 135 (1993).

\bibitem{CleymansRedlich1998}
J. Cleymans, K. Redlich, Phys. Rev. Lett. {\bf 81}, 5284 (1998).

\bibitem{CleymansRedlich1999}
J. Cleymans, K. Redlich, Phys. Rev. C {\bf 60}, 054908 (1999).

\bibitem{Becattini2001}
F.~Becattini, J.~Cleymans, A.~Keranen, E.~Suhonen, and K.~Redlich,
Phys. Rev. C {\bf 64}, 024901 (2001).

\bibitem{Becattini2004}
F.~Becattini, M.~Gazdzicki, A.~Keranen, J.~Manninen, and R.~Stock, Phys. Rev. C {\bf 69}, 024905 (2004).

\bibitem{ABS2006}
A. Andronic, P. Braun-Munzinger, and J. Stachel, Nucl. Phys. A {\bf 772}, 167 (2006).

\bibitem{ABS2009}
A. Andronic, P. Braun-Munzinger, and J. Stachel, Phys. Lett. B {\bf 673}, 142 (2009).

\bibitem{DMB}
R. Dashen, S.-K. Ma, and H. J. Bernstein, Phys. Rev. {\bf 187}, 345
(1969).

\bibitem{EV-latt-1} 
A. Andronic, P. Braun-Munzinger, J. Stachel, and M. Winn, 
Phys. Lett. B {\bf 718}, 80 (2012).  

\bibitem{EV-latt-2}
V.~Vovchenko, D.~V.~Anchishkin, and M.~I.~Gorenstein,
Phys.\ Rev.\ C {\bf 91}, 024905 (2015).

\bibitem{EV-latt-3} 
M. Albright, J. Kapusta, and C. Young, Phys. Rev. C {\bf 90}, 024915 (2014);
Phys. Rev. C {\bf 92}, 044904 (2015).


\bibitem{VS2015}
V. Vovchenko and H. Stoecker,
arXiv:1512.08046 [hep-ph], J. Phys. G, in print.



\bibitem{HagedornEV1}
R. Hagedorn and J. Rafelski, Phys. Lett. B {\bf 97}, 136 (1980);
R. Hagedorn, Z. Phys. C {\bf 17}, 265 (1983).

\bibitem{Gorenstein1981}
M. I. Gorenstein, V. K. Petrov, and G. M. Zinovjev,
Phys. Lett. B {\bf 106}, 327 (1981).

\bibitem{Kapusta1983} 
  J.~I.~Kapusta and K.~A.~Olive,
  Nucl.\ Phys.\ A {\bf 408}, 478 (1983).

\bibitem{Rischke1991}
D. H. Rischke, M. I. Gorenstein, H. St\"ocker, and W. Greiner, Z. Phys. C {\bf 51}, 485 (1991). 

\bibitem{Yen1997} 
  G.~D.~Yen, M.~I.~Gorenstein, W.~Greiner, and S.~N.~Yang,
  Phys.\ Rev.\ C {\bf 56}, 2210 (1997).

\bibitem{LL} 
L. D. Landau and E. M. Lifshitz, Statistical Physics (Oxford: Pergamon) 1975.

\bibitem{GKK} 
  M.~I.~Gorenstein, A.~P.~Kostyuk and Y.~D.~Krivenko,
  J.\ Phys.\ G {\bf 25}, L75 (1999).
  
\bibitem{BugaevEV} 
  K.~A.~Bugaev, D.~R.~Oliinychenko, A.~S.~Sorin, and G.~M.~Zinovjev,
  Eur.\ Phys.\ J.\ A {\bf 49}, 30 (2013).

\bibitem{Broyden}
  C.~G.~Broyden, 
  Mathemathics of Computation {\bf 19}, 577 (1965).

\bibitem{pdg}
  K.~A.~Olive {\it et al.}  [Particle Data Group Collaboration],
  Chin.\ Phys.\ C {\bf 38}, 090001 (2014).

\bibitem{Broniowski:2015oha}
  W.~Broniowski, F.~Giacosa, and V.~Begun,
  Phys.\ Rev.\ C {\bf 92}, 034905 (2015).

\bibitem{Pelaez:2015qba}
  J.~R.~Pelaez,
  Phys.\ Rept.\  {\bf 658}, 1 (2016).
  
  \bibitem{PHS1999} 
  P.~Braun-Munzinger, I.~Heppe, and J.~Stachel,
  Phys.\ Lett.\ B {\bf 465}, 15 (1999).
  
\bibitem{Cleymans2006} 
  J.~Cleymans, H.~Oeschler, K.~Redlich, and S.~Wheaton,
  Phys.\ Rev.\ C {\bf 73}, 034905 (2006).

\bibitem{Begun2013}
  V.~V.~Begun, M.~Gazdzicki, and M.~I.~Gorenstein,
  Phys.\ Rev.\ C {\bf 88}, 024902 (2013).

\bibitem{mf-2014}
D. Anchishkin, V. Vovchenko,
J. Phys. G {\bf 42}, 105102 (2015).


\bibitem{Gorenstein1999}
G.~D.~Yen and M.~I.~Gorenstein,
  Phys.\ Rev.\ C {\bf 59}, 2788 (1999).

\bibitem{NoronhaHostler2009} 
  J.~Noronha-Hostler, H.~Ahmad, J.~Noronha, and C.~Greiner,
  Phys.\ Rev.\ C {\bf 82}, 024913 (2010).

\bibitem{Becattini2006}
F. Becattini, J. Manninen, and M. Ga\'zdzicki, Phys. Rev. C {\bf
73}, 044905 (2006).


\bibitem{NM-VDW}
  V.~Vovchenko, D.~V.~Anchishkin, and M.~I.~Gorenstein,
  Phys.\ Rev.\ C {\bf 91}, 064314 (2015).

\bibitem{NA49data-1}
S.V.~Afanasiev {\it et al.} [NA49 collaboration], Phys. Rev. C {\bf 66}, 054902 (2002);
  C.~Alt {\it et al.} [NA49 Collaboration],
  Phys.\ Rev.\ C {\bf 73}, 044910 (2006);
C.~Alt {\it et al.} [NA49 collaboration], Phys. Rev. C {\bf 77}, 024903 (2008).

\bibitem{NA49data-2}
  C.~Alt {\it et al.} [NA49 Collaboration],
  Phys.\ Rev.\ C {\bf 78}, 034918 (2008);
%
  C.~Alt {\it et al.} [NA49 Collaboration],
  Phys.\ Rev.\ C {\bf 78}, 044907 (2008);
%
  C.~Alt {\it et al.} [NA49 Collaboration],
  Phys.\ Rev.\ Lett.\  {\bf 94}, 192301 (2005);
%
%
  T.~Anticic {\it et al.} [NA49 Collaboration],
  Phys.\ Rev.\ C {\bf 85}, 044913 (2012);
%
  V.~Friese [NA49 Collaboration],
  Nucl.\ Phys.\ A {\bf 698}, 487 (2002).
%
\bibitem{NA49data-3}
The tabulated hadron yield data of the NA49 collaboration is available at
https://edms.cern.ch/document/1075059.

\bibitem{STARdata}
  J.~Adams {\it et al.} [STAR Collaboration],
  Phys.\ Rev.\ Lett.\  {\bf 92}, 112301 (2004);
  Phys.\ Rev.\ Lett.\  {\bf 98}, 062301 (2007);
  B.~I.~Abelev {\it et al.} [STAR Collaboration],
  Phys.\ Rev.\ Lett.\  {\bf 99}, 112301 (2007).

\bibitem{BecattiniSTAR}
  J.~Manninen and F.~Becattini,
  Phys.\ Rev.\ C {\bf 78}, 054901 (2008).
  

\bibitem{ALICEdata}
  B.~Abelev {\it et al.} [ALICE Collaboration],
  Phys.\ Rev.\ C {\bf 88}, 044910 (2013);
  B.~B.~Abelev {\it et al.} [ALICE Collaboration],
  Phys.\ Rev.\ Lett.\  {\bf 111}, 222301 (2013);
  B.~B.~Abelev {\it et al.} [ALICE Collaboration],
  Phys.\ Lett.\ B {\bf 728}, 216 (2014);
  B.~B.~Abelev {\it et al.} [ALICE Collaboration],
  Phys.\ Rev.\ C {\bf 91}, 024609 (2015).

\bibitem{Becattini2014}
F. Becattini, E. Grossi, M. Bleicher, J. Steinheimer, and  R.~Stock, Phys. Rev. C {\bf 90}, 054907 (2014).

\bibitem{Letessier2005} 
  J.~Letessier and J.~Rafelski,
  Eur.\ Phys.\ J.\ A {\bf 35}, 221 (2008).

\bibitem{MHBQ} 
  J.~Rafelski,
  Eur.\ Phys.\ J.\ A {\bf 51}, 114 (2015).

\bibitem{VBG2015}
V.~Vovchenko, V.~V.~Begun, and M.~I.~Gorenstein,
  Phys.\ Rev.\ C {\bf 93}, 064906 (2016).

\bibitem{ContFrz1}
  S.~A.~Bass, A.~Dumitru, M.~Bleicher, L.~Bravina, E.~Zabrodin, H.~Stoecker, and W.~Greiner,
  Phys.\ Rev.\ C {\bf 60}, 021902 (1999).
\bibitem{ContFrz2}
L.~V.~Bravina, I.~N.~Mishustin, J.~P.~Bondorf, A.~Faessler, and E.~E.~Zabrodin,
  Phys.\ Rev.\ C {\bf 60}, 044905 (1999).

\bibitem{ZalewskiRedlich} 
  K.~Redlich and K.~Zalewski,
  Phys.\ Rev.\ C {\bf 93}, 014910 (2016).

\bibitem{Aoki:2006we} 
  Y.~Aoki, G.~Endrodi, Z.~Fodor, S.~D.~Katz, and K.~K.~Szabo,
  Nature {\bf 443}, 675 (2006).
  
\bibitem{lQCD} 
  S.~Borsanyi, Z.~Fodor, C.~Hoelbling, S.~D.~Katz, S.~Krieg, and K.~K.~Szabo,
  Phys.\ Lett.\ B {\bf 730}, 99 (2014).

\bibitem{Bazavov:2014pvz} 
  A.~Bazavov {\it et al.} [HotQCD Collaboration],
  Phys.\ Rev.\ D {\bf 90}, 094503 (2014).

\bibitem{Mukherjee:2015mxc} 
  S.~Mukherjee, P.~Petreczky, and S.~Sharma,
  Phys.\ Rev.\ D {\bf 93}, 014502 (2016).

\bibitem{Biro:2014sfa} 
  T.~S.~Biro and A.~Jakovac,
  Phys.\ Rev.\ D {\bf 90}, 094029 (2014).

\bibitem{Jakovac:2013iua} 
  A.~Jakovac,
  Phys.\ Rev.\ D {\bf 88}, 065012 (2013).

\bibitem{Stoecker:1980uk} 
  H.~Stoecker, G.~Graebner, J.~A.~Maruhn, and W.~Greiner,
  Phys.\ Lett.\  {\bf B95}, 192 (1980).

\bibitem{Becattini2013} 
  F.~Becattini, M.~Bleicher, T.~Kollegger, T.~Schuster, J.~Steinheimer, and R.~Stock,
  Phys.\ Rev.\ Lett.\  {\bf 111}, 082302 (2013).

\bibitem{Becattini:2016xct} 
  F.~Becattini, J.~Steinheimer, R.~Stock, and M.~Bleicher,
  Phys.\ Lett.\ B {\bf 764}, 241 (2017)

\bibitem{Satarov:2016peb} 
  L.~M.~Satarov, V.~Vovchenko, P.~Alba, M.~I.~Gorenstein, and H.~Stoecker,
  Phys.\ Rev.\ C {\bf 95}, 024902 (2017).
  
\end{thebibliography}
\end{document}